\documentclass[%
 aip,
 jcp,%
 amsmath,
 amssymb,
 preprint,%
]{revtex4-2}

\usepackage{threeparttable,array,booktabs,calc}
\usepackage{graphicx}
\usepackage{dcolumn}
\usepackage{bm}

\usepackage[version=4]{mhchem}

\usepackage{hyperref}

\usepackage{epstopdf} 

\usepackage[final]{changes} 

\newcommand{\unit}[1]{\,\mathrm{#1}}

\newcommand{\uunit}[1]{\,\mu\mathrm{#1}} 

\begin{document}

\preprint{SpinPolPaper-1}

\title{Spin-state-controlled chemi-ionization reactions between metastable helium atoms and ground-state lithium atoms}

\author{Tobias Sixt}
\affiliation{Institute of Physics, University of Freiburg, Hermann-Herder-Str. 3, 79104 Freiburg, Germany}
\author{Frank Stienkemeier}
\affiliation{Institute of Physics, University of Freiburg, Hermann-Herder-Str. 3, 79104 Freiburg, Germany}
\author{Katrin Dulitz}
\email{katrin.dulitz@physik.uni-freiburg.de.}
\affiliation{Institute of Physics, University of Freiburg, Hermann-Herder-Str. 3, 79104 Freiburg, Germany}

\date{\today}

\begin{abstract}
We demonstrate the control of $^4$He(2$^3$S$_1$)-$^7$Li(2$^2$S$_{1/2}$) chemi-ionization reactions by all-optical electron-spin-state preparation of both atomic species prior to the collision process. Our results demonstrate that chemi-ionization is strongly suppressed (enhanced) for non-spin-conserving (spin-conserving) collisions at thermal energies. These findings are in good agreement with a model based on spin angular momentum coupling of the prepared atomic states to the quasi-molecular states. Small deviations from the model indicate the contribution of the $^4\Sigma^+$ channel to the reaction rate which is in violation of spin conservation.
\end{abstract}
\pacs{Valid PACS appear here}
%
\maketitle

\section{\label{sec:introduction}Introduction}
\noindent Trapped ultracold mixtures of different atomic species are the starting point for the production of dense samples of ultracold heteronuclear molecules which may feature long-range and anisotropic interactions \cite{Molony2014, Takekoshi2014, Park2015, Guo2016, Voges2020}. Such interactions allow for new physics and chemistry studies in a regime which is purely dominated by quantum effects \replaced{\cite{Carr2009,Hu2021}}{\cite{Carr2009}}. To achieve the efficient co-trapping of ultracold alkali atoms and metastable rare-gas atoms, chemi-ionization reactions must be efficiently suppressed.

Chemi-ionization occurs when an excited, long-lived (``metastable'') atom or molecule A$^*$ reacts with an atom or molecule B whose ionization energy is lower than the internal energy of A$^*$:
\begin{align}
	\mathrm{A}^* + \mathrm{B} \rightarrow [\mathrm{AB}]^* \rightarrow
	\begin{cases}
		\mathrm{A} + \mathrm{B}^+ + \mathrm{e}^- \,\,(\mathrm{PI})\\
		\mathrm{AB}^+ + \mathrm{e}^- \,\,(\mathrm{AI})
	\end{cases}
	\label{eq:PIAI}
\end{align}
In Eq.~\eqref{eq:PIAI}, the product pathways are referred to as Penning ionization (PI) and associative ionization (AI), respectively. \added{If B is a molecule, rearrangement and dissociative ionization are possible product channels as well \cite{Yencha1978,Haberland1981}.}

The results of recent experiments have revealed that quantum effects, such as orbiting resonances, as well as stereodynamic effects strongly affect chemi-ionization rates \cite{Henson2012,Jankunas2015b,Gordon2017,Zou2018,Gordon2018,Paliwal2021}.
%
In ultracold trapped samples of $^4$He(2$^3$S$_1$) and $^3$He(2$^3$S$_1$), it has been shown theoretically and experimentally that rapid trap losses arising from chemi-ionization can be suppressed by several orders of magnitude using spin-state preparation \cite{Shlyapnikov1994, Herschbach2000,McNamara2006}. This suppression is observed when the atoms are prepared in electron-spin-stretched states. In this case, product formation by chemi-ionization is forbidden as a result of electron-spin conservation (Wigner's spin conservation rule \cite{Wigner1928}), since the total electron spin of the prepared reaction partners in the entrance channel is different from the total electron spin of the reaction products in the exit channel.

Spin suppression was also observed in chemi-ionization studies of spin-polarized, co-trapped heteronuclear mixtures of $^4$He(2$^3$S$_1$) and $^{87}$Rb($5^2$S$_{1/2}$) at ultracold temperatures \cite{Byron2010, Knoop2014, Flores2016, Flores2017}. Since the exit channel in a He$^*$-alkali-atom system is of $^2\Sigma^+$ character, spin conservation implies that the entrance channels of $^2\Sigma^+$ symmetry ($^4\Sigma^+$ symmetry) are reactive (unreactive). These assumptions were also made in previous studies of He$^*$-alkali-atom systems at low and at thermal collision energies 
\cite{Ruf1987, Siska1993, Knoop2014, Kedziera2015}.

For homonuclear reactions of the heavy rare-gas atoms, it was found that spin-orbit effects lift the spin-conservation restriction (see Ref.\ \cite{Vassen2012} and references therein). \textit{Ab-initio} calculations of reactive collisions involving metastable atoms are demanding, because they require the calculation of two-center two-electron integrals \cite{OpdeBeek1997a}. \replaced{Recent advances in the theoretical modeling of chemi-ionization also enable the full description of state-to-state controlled reaction stereodynamics without the difficulties connected to \textit{ab-initio} calculations \cite{Falcinelli2020,Falcinelli2020a} and may be extended to explain spin-state-dependent effects as well. A comparison in between theoretical and experimental results is always desirable in order to identify effects that have not been accounted for in theory work.}{Therefore, these properties are best determined experimentally.}

In this article, we present the first experimental study of chemi-ionization reactions between spin-state-prepared $^4$He atoms in the metastable $2^3$S$_1$ level (referred to as He$^*$ hereafter) and ultracold $^7$Li atoms in the $2^2$S$_{1/2}$ electronic ground state (referred to as Li hereafter). This study at thermal collision energies is a first step towards the co-trapping of He$^*$ and Li at ultracold temperatures. For the experiments detailed here, we have combined a supersonic-beam source for He$^*$ with a magneto-optical trap (MOT) for Li. In order to distinguish in between the contributions of He($2^3$S$_1$) and He($2^1$S$_0$) to the reaction rate, we deplete the population in the $2^1$S$_0$ level using an optical-excitation scheme developed by us \cite{Guan2019}. Prior to studying the collision process itself using time-of-flight detection of the product ions, we use laser-optical pumping to prepare both He$^*$ and Li in specific magnetic sublevels.
\section{\label{sec:experimental}Experimental}
\subsection{General setup}
\noindent Most parts of the experimental setup have already been described elsewhere \cite{Grzesiak2019, Guan2019, Dulitz2020a, Sixt2021}. In the following, we thus briefly provide details about the most important parts of the setup and about those features which have specifically been implemented for this study.

\begin{figure}[!htbp]
	\includegraphics{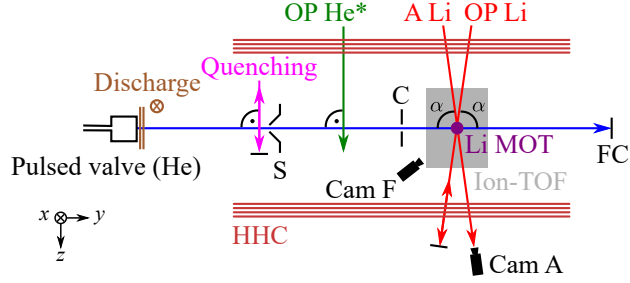}
	\caption{\label{fig:setup} Sketch of the experimental setup (see main text), not to scale. The velocity axis of the pulsed He atomic beam is indicated as a blue horizontal arrow. A homogeneous magnetic field in the $z$ direction is produced by a pair of Helmholtz coils (HHC) which is located along the $xy$-plane. The laser beams for the optical pumping of Li (OP Li) and the absorption imaging of Li (A Li) are tilted at an angle of $\alpha' = 90^\circ-\alpha = 4^\circ$ with respect to the $z$ axis. Other abbreviations: \replaced{S}{S$\,\ldots\,$s}kimmer \added{(S)}, \deleted{C$\,\ldots\,$}collimating aperture \added{(C)}, \deleted{Cam F$\,\ldots\,$}CCD camera used to monitor the fluorescence of excited Li atoms \added{(Cam F)}, \deleted{Cam A$\,\ldots\,$}CCD camera used for the absorption imaging of Li \added{(Cam A)}, \deleted{Ion-TOF$\,\ldots\,$}ion-time-of-flight detector \added{(Ion-TOF)}, \deleted{FC$\,\ldots\,$}Faraday-cup detector \added{(FC)}, \deleted{OP He$^*$$\,\ldots\,$} laser beam for the optical pumping of He$^*$ \added{(OP He$^*$)}, \deleted{Li MOT$\,\ldots\,$}magneto-optical trap for ultracold Li atoms \added{(Li MOT)}.}
\end{figure}
A sketch of the experimental setup is shown in Fig.\ \ref{fig:setup}. A pulsed beam of ground-state $^4$He atoms is produced in a supersonic expansion of $^4$He gas ($10\unit{bar}$ backing pressure) through a pulsed valve ($30 \uunit{s}$ pulse duration) into the vacuum. The pulsed valve is mounted to a Gifford-McMahon cryocooler which allows for easy adjustments of the valve temperature $T_{\mathrm{valve}}$. A\deleted{n} small fraction of the atoms ($\approx 10^{-4}$) is excited to the metastable $2^3$S$_1$ and $2^1$S$_0$ levels by electron impact excitation in an electron-seeded discharge \cite{Grzesiak2019}. After discharge excitation, more than $99\,\%$ of the He atoms in the $2^1$S$_0$ level are quenched to the $1^1$S$_0$ electronic ground state by laser excitation via the $4^1$P$_1 \leftarrow 2^1$S$_0$ transition at $\lambda = 397 \unit{nm}$ and subsequent rapid decay \cite{Guan2019}. After collimation by a $1\unit{mm}$-diameter skimmer, the atomic beam enters another vacuum chamber, in which the spin orientation of He atoms in the 2$^3$S$_1$ level (He$^*$) is prepared using laser-optical pumping (OP). For this, the $2^3$P$_2 \leftarrow 2^3$S$_1$ transition in He is excited using circularly polarized laser light at $\lambda = 1083 \unit{nm}$. To analyze the relative magnetic-sublevel populations in He$^*$, the fluorescence emitted by the atoms during optical pumping is collected using a lens system in combination with a polarizer plate and a photo-diode detector. More details about the optical pumping scheme for He$^*$ and about the fluorescence detection system are discussed in Ref.\ \cite{Sixt2021}.

After optical pumping, the atomic beam passes an additional collimating aperture ($3\unit{mm}$ diameter) and enters the reaction region. Here, an ultracold cloud of $\approx 6 \cdot 10^7$ $^7$Li atoms in the 2$^2$S$_{1/2}$ level, confined in a magneto-optical trap (MOT), serves as a stationary scattering target. The Li atoms are loaded into the MOT from a continuous atomic beam of Li atoms produced inside a heated oven. An ion-time-of-flight (TOF) detector with a channel electron multiplier (CEM) in counting mode, built around the reaction region, is used to detect the ions produced by chemi-ionization. Owing to the low kinetic energy of the Li atoms, the He$^*$-Li collision energy is solely dependent on the forward velocity of the He$^*$ atoms. The most probable forward velocity and the intensity of the pulsed beam of He$^*$ are monitored using a gold-coated Faraday cup detector (FC) which is placed further downstream from the reaction region. In the experiments described here, the most probable forward velocity of the pulsed beam of He$^*$ is varied in between $v = 1824\unit{m/s}$ at $T_\text{valve} = 300\unit{K}$ and $v = 853\unit{m/s}$ at $T_\text{valve} = 50\unit{K}$. 

A magnetic bias field of $B_z \approx 3\unit{G}$, produced by a pair of Helmholtz coils, defines the quantization axis for the He$^*$ atoms and for the Li atoms during optical pumping as well as during the reaction process.
\subsection{Preparation and detection of the relative magnetic-sublevel populations in Li}
\noindent The ultracold Li atoms are spin-polarized prior to the collision process using laser-optical pumping (OP) with circularly polarized laser light following the ideas given by Guillot et al. \cite{Gillot2013}.

Two spatially overlapped laser beams whose frequencies are resonant with the hyperfine transitions $F' = 2 \leftarrow F'' = 2$ and $F' = 2 \leftarrow F'' = 1$ within the $2^2$P$_{1/2} \leftarrow 2^2$S$_{1/2}$ (D$_1$) transition at $\lambda = 671\unit{nm}$, respectively, are generated using a home-built diode laser system. Here, a double prime ($''$) is used to denote the lower-lying level, whereas a single prime ($'$) is used to label the higher-lying level. Each laser beam has a power of $P = 4.4\unit{mW}$ and a Gaussian waist size of $w = 4.8\unit{mm}$. For each laser beam, an acousto-optic modulator is used to switch the laser power on and off within a time period of $\approx 100\unit{ns}$. The amount of left- and right-handed circular polarization of the laser light is adjusted using a rotatable quarter-wave plate before the laser beams enter the vacuum chamber through an anti-reflection-coated entrance window. For technical reasons, the angle of incidence of the laser light on the Li atom cloud is tilted by $\alpha' = 4^\circ$ with respect to $B_z$.

The laser-Li interaction results in a population transfer to the magnetic sublevel $F'' = 2$, $M_{F''} = \pm 2$ of the $2^2$S$_{1/2}$ electronic ground state if the light is $\sigma^\pm$-polarized. Atoms in the respective sublevel do not interact with the circularly polarized laser light anymore (dark state), because there is no excited sublevel that the laser could couple to. However, there are several factors, such as the polarization purity of the OP light or the angle in between the laser beam for OP and the quantization axis, which limit the population transfer efficiency. Therefore, an absorption laser system is used as a diagnostic tool for the determination of the relative magnetic-sublevel populations in the electronic ground state. The absorption laser light is generated by a grating-stabilized commercial diode laser (Toptica DL100) at $\lambda = 671\unit{nm}$. A frequency-offset locking scheme \cite{Schuenemann1999} is used to stabilize the frequency of the absorption laser relative to the frequency of the MOT master laser. The latter is locked to the cross-over signal of the $F' = 3 \leftarrow F'' = 2$ and $F' = 2 \leftarrow F'' = 1$ hyperfine transitions within the $2^2$P$_{3/2} \leftarrow 2^2$S$_{1/2}$ (D$_2$) transition of Li. The absorption laser beam is collimated to $w = 4.8\unit{mm}$ and its polarization is adjusted using a rotatable quarter-wave plate. The angle of incidence of the laser beam on the Li atom cloud is tilted by $\alpha' = 4^\circ$ with respect to $B_z$. The power of the absorption laser beam is low ($P = 15\uunit{W}$) so that a saturation of the atomic transition during the absorption measurement (saturation parameter $s = I/I_\text{sat} = 0.016$ with a saturation intensity of $I_\text{sat} = 2.56\unit{mW/cm^2}$ for the D$_2$ transition in $^7$Li \cite{Metcalf1999}) is avoided. The transmitted laser light is recorded using a CCD camera (Cam A). According to the Beer–Lambert law, the optical density (OD) of the Li cloud at each laser frequency detuning from resonance $\Delta$ is given by 
\begin{align}
	\mathrm{OD}(\Delta) = - \ln \left( \frac{I(\Delta)}{I_0} \right),
\end{align}
where $I_0$ is the incident laser intensity and $I(\Delta)$ is the transmitted laser intensity at a specific laser detuning $\Delta$.
The intensities are deduced from the sum of all pixels of the CCD camera. The OD is normalized to the number of Li atoms obtained by fluorescence imaging using an additional CCD camera (Cam F). This normalization procedure compensates for any OD variations arising from a change of the total Li atom number.
\subsection{Timing sequences and experimental protocol}
\begin{figure}[!htbp]
	\includegraphics{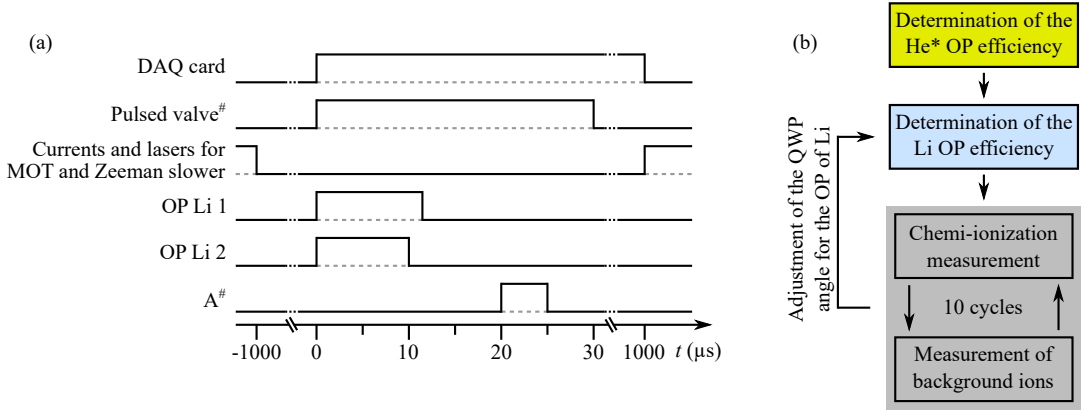}
	\caption{\label{fig:sequence} 
	(a) Illustration of the timing sequence used for each variation of the experimental protocol shown in (b). The three different subroutines of the experimental protocol are color-coded (yellow, blue and gray shadings). OP Li 1 (OP Li 2) denotes the laser beam for the optical pumping of Li whose frequency is resonant with the $F' = 2 \leftarrow F'' = 1$ ($F' = 2 \leftarrow F'' = 2$) hyperfine transition within the D$_1$ transition. Note that either the pulsed valve or the laser for Li absorption imaging (A) are active at the same time (denoted with a \# sign). Other abbreviations: \replaced{D}{DAQ card$\ldots$d}ata acquisition card \added{(DAQ card)}, \deleted{OP$\ldots$}optical pumping \added{(OP)}, \deleted{QWP$\ldots$}quarter-wave plate \added{(QWP)}.}
\end{figure}
\noindent The measurements are done at three different valve temperatures and at a repetition rate of $5\unit{Hz}$ using the timing sequences and the experimental protocol shown in Figs.\ \ref{fig:sequence} (a) and (b), respectively. The experimental protocol is subdivided into three subroutines (illustrated by the three different color shadings in Fig.\ \ref{fig:sequence} (b)), each being performed multiple times. For each subroutine, $t = 0\uunit{s}$ is given by the simultaneous triggering of the pulsed valve and the data acquisition card. The currents used to generate the magnetic fields for the MOT and for the Zeeman slower are switched off $1\unit{ms}$ before that, so that their magnetic fields do not produce magnetic-field inhomogeneities during the laser-optical pumping and during the collision experiments, respectively. The laser beams used for Zeeman slowing and for the generation of the MOT are switched off at the same time.

The three subroutines are as follows:
\begin{enumerate}
	\item The relative magnetic-sublevel populations in He$^*$ are determined following the procedure described in Ref.\ \cite{Sixt2021}. In short, this involves an analysis of the polarization state of the fluorescence light emitted by the He atoms while they are being optically pumped. To achieve the maximum transfer efficiency to the $M_{J''} = + 1$ ($M_{J''} = - 1$) sublevel of He$^*$, the laser polarization for optical pumping is set to $\sigma^+$($\sigma^-$) and a laser power of $P = 200\unit{mW}$ is used \cite{Sixt2021}.
	\item The relative magnetic-sublevel populations in the Li $2^2$S$_{1/2}, F = 2$ hyperfine level are prepared using optical pumping and measured using absorption imaging. For optical pumping of the $F' = 2 \leftarrow F'' = 1$ ($F' = 2 \leftarrow F'' = 2$) hyperfine transition of the D$_1$ transition at $\lambda = 671\unit{nm}$, the resonant laser beam is turned on at $t = 0\uunit{s}$ and switched off after a time interval of $10\uunit{s}$ ($11\uunit{s}$). In this way, it is ensured that the $2^2$S$_{1/2}, F'' = 1$ level is fully depopulated. During the time interval from $t = 20\uunit{s}$ to $t = 25\uunit{s}$, the Li atoms are exposed to the absorption laser light, and the corresponding absorption image is recorded using Cam A. The absorption laser frequency is scanned in a stepwise manner ($\approx 3\unit{MHz}$ step size) around the $F' = 3, 2, 1 \leftarrow F'' = 2$ hyperfine transitions within the D$_2$ transition. The frequency scan is done twice, first using $\sigma^+$-polarized absorption laser light and then using $\sigma^-$-polarized absorption laser light.
	\item The chemi-ionization measurements are done using the relative He$^*$ and Li magnetic-sublevel populations prepared in subroutines 1 and 2. Li optical pumping is turned off for every second shot of the pulsed valve in order to take a reference chemi-ionization measurement for non-spin-polarized Li atoms. In addition to that, a background measurement without the presence of Li atoms in the reaction region is done to distinguish between the ions produced by He$^*$-Li chemi-ionization and by other chemi-ionization reactions involving He$^*$. For each Li magnetic-sublevel population distribution prepared in subroutine 2, ten He$^*$-Li and ten background chemi-ionization measurements are done in toggle mode.
\end{enumerate} 
Once subroutine 3 is completed, the polarization of the lasers for Li optical pumping is changed and subroutines 2 and 3 are repeated. An incremental change in between $\sigma^+$ and $\sigma^-$ polarization is achieved by changing the angle of the quarter-wave plate for Li optical pumping in five equal steps within a total range of $90^\circ$. In this way, the spin orientation of Li is scanned from aligned to anti-aligned with respect to the spin orientation of He$^*$ .
%
\section{\label{sec:theory}\added{Modelling}\deleted{Theoretical description} of \deleted{the}chemi-ionization rate ratios}
\begin{figure}[!htbp]
	\includegraphics{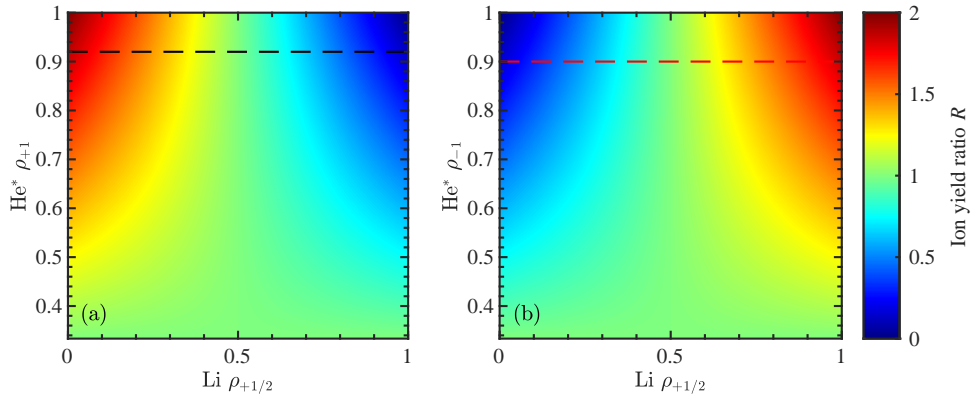}
	\caption{\label{fig:heatmap} Modeled ion yield ratios obtained from Eq.~\eqref{eq:modelfit} using $k_4 = 0$ and $\rho^\text{(eq)}_{+1/2} = \rho^\text{(eq)}_{-1/2} = 0.5$. The relative spin-sublevel populations in Li are shown over the full range, assuming that $\rho_{-1/2} = 1 - \rho_{+1/2}$, while the relative spin-sublevel populations in He$^*$ are shown for (a) $M_{J''} = M_S = +1$ and (b) $M_{J''} = M_S = -1$. In the case of He$^*$, we assume that the remaining population is equally distributed over the remaining spin-sublevels, i.e. $\rho_0 = \rho_{-1} = (1 - \rho_{+1})/2$ for (a) and $\rho_0 = \rho_{+1} = (1 - \rho_{-1})/2$ for (b). Therefore, $\rho_{+1} = \rho_{-1} = 1/3$ corresponds to an equal population of the spin-sublevels of He$^*$. The dashed black (red) line indicates the modeled ion yield ratios for the experimentally prepared relative spin-sublevel population $\rho_{+1}$ ($\rho_{-1}$) in He$^*$.}
\end{figure}

\noindent \added{
	For He$^*$-Li chemi-ionization, the collision complex can be formed within the $^2 \Sigma^+$ or the $^4 \Sigma^+$ entrance channel, respectively. This collision complex autoionizes by coupling to the continuum of states of the $(\mathrm{HeLi})^+ +\mathrm{e}^-$ system. The (HeLi)$^+$ $^2\Sigma^+$ exit channel is the lower boundary of this continuum. Here, a simplified, heuristic model is used in which the exact shapes of the entrance and exit channel potentials are neglected and a fixed collision energy is used. In this model, the initial states formed within the $^2\Sigma^+$ ($^4\Sigma^+$) entrance channel are assigned the chemi-ionization rate constant $k_a$ with $a = 2$ ($a = 4$).} \replaced{Consequently, t}{T}he ion production rate for He$^*$-Li chemi-ionization via a specific \replaced{entrance channel}{quasi-molecular potential} $a$ is given by 
\begin{align}
	[\dot{I}_a] = k_a [\mathrm{He}^*_a] [\mathrm{Li}_a],
\end{align}
where \deleted{$k_a$ is the chemi-ionization rate constant in the $a$th quasi-molecular potential, with $a = 2$ ($a = 4$) for the $^2\Sigma^+$ ($^4\Sigma^+$) quasi-molecular potential, respectively.} \replaced{t}{T}he corresponding He$^*$ and Li densities are denoted as $[\mathrm{He^*}_a]$ and $[\mathrm{Li}_a]$. However, in our experiment, only the quantum states of the two atomic species are prepared prior to the collision process which corresponds to the asymptotic part of the \replaced{entrance channel}{quasi-molecular potential}. Thus, we introduce a constant $C_{a,b,c}$ which couples the prepared atomic states $b$ and $c$ to the \replaced{entrance channel}{quasi-molecular potential} $a$, so that
\begin{align}
	[\dot{I}_a] = k_a \cdot C_{a,b,c} \cdot [\mathrm{He}^*_b] \cdot [\mathrm{Li}_c].
	\label{eq:ionrate}
\end{align}
The constant $C_{a,b,c}$ is obtained by angular momentum coupling of the prepared atomic spin sublevels to the total spin sublevels within the quasi-molecular \replaced{states}{potentials}. Consequently, the total ion production rate is given by considering all possible initial atomic state combinations and their resulting contributions to the quasi-molecular states:
\begin{align}
	[\dot{I}]_\text{tot} = \sum_{a,b,c} k_a \cdot C_{a,b,c} \cdot [ \mathrm{He}^*_b ] \cdot [ \mathrm{Li}_c ].
\end{align}
Since the determination of absolute densities suffers from large systematic errors, we focus on the measurement of relative atomic densities and ion yield ratios $R$. Finally, all experimentally observed ion yield ratios are expressed as
\begin{align}
	R &= \frac{[I]^{\text{(prep)}}_\text{tot}}{[I]^{\text{(eq)}}_\text{tot}} \\
	&= \frac{\sum_{b,c} \rho^{\text{(prep)}}_b \cdot \rho^{\text{(prep)}}_c \cdot \left( C_{2,b,c} + k_4/k_2 \cdot  C_{4,b,c}\right)}{\sum_{b,c}  \rho^{\text{(prep)}}_b \cdot \rho^{\text{(eq)}}_c \cdot \left( C_{2,b,c} + k_4/k_2 \cdot  C_{4,b,c} \right)},
	\label{eq:modelfit}
\end{align}
where the ion yields are obtained from an integration of the ion count rates over time. The rate constants $k_2$ and $k_4$ of the $^2 \Sigma^+$ and $^4 \Sigma^+$ \replaced{entrance channels}{quasi-molecular potentials}, respectively, are re-arranged to the rate constant ratio $k_4/k_2$. Thus, we account for any violation of spin conservation by assuming a finite contribution of the $^4\Sigma^+$ \replaced{entrance channel}{quasi-molecular potential} to the chemi-ionization rate. 
In the equation above, $\rho^{\text{(prep)}}_b$ and $\rho^{\text{(prep)}}_c$ are the relative spin-sublevel populations in He$^*$ and Li, respectively, which are prepared using optical pumping. \added{For normalization purposes, t}\deleted{T}he relative spin-sublevel population in Li \added{is also prepared}\deleted{under} \added{in} equilibrium\deleted {conditions is}\added{, }denoted as $\rho^{\text{(eq)}}_c$\added{ in Eq. \ref{eq:modelfit}}. The value for $\rho^{\text{(eq)}}_c$ is obtained from optical pumping with an equal mixture of $\sigma^+$- and $\sigma^-$-polarized light. Fig.\ \ref{fig:heatmap} shows the dependence of the modeled ion yield ratio as a function of the relative spin-sublevel populations in Li and He$^*$. In Fig.\ \ref{fig:heatmap} (a) (Fig.\ \ref{fig:heatmap} (b)), this dependency is shown as a function of the relative spin-sublevel population in He$^*$ with $M_{J''} = M_S = +1$ ($M_{J''} = M_S = -1$).
\section{\label{sec:results}Results and Discussion}
\begin{figure}[!htbp]
	\includegraphics{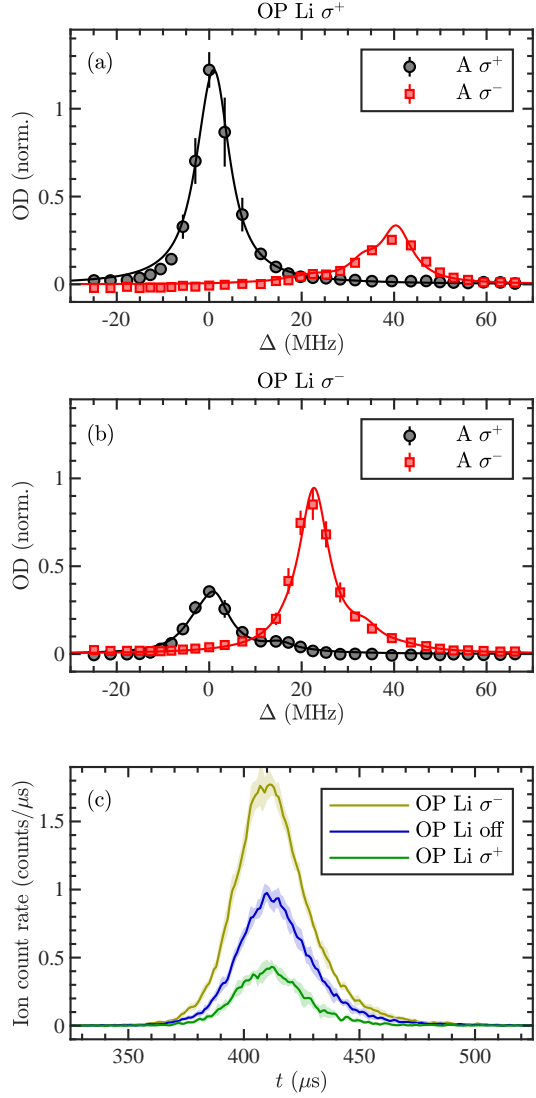}
	\caption{\label{fig:ionsSpectrum} (a), (b) Li absorption spectra after optical pumping of the atoms using $\sigma^+$-polarized and $\sigma^-$-polarized light, respectively. Black circles (red squares) correspond to measured absorption spectra in which the absorption laser is $\sigma^+$-polarized ($\sigma^-$-polarized). The error bars represent the statistical fluctuations of 20 measurements. Fitted absorption spectra obtained for each absorption laser polarization are shown as solid lines. (c)  Background-corrected ion count rates from He$^*$-Li chemi-ionization at a collision energy of $E_\text{coll} = 43.9\unit{meV}$ as a function of time with respect to $t = 0\uunit{s}$ using $\sigma^-$-polarized light for Li optical pumping (yellow curve), $\sigma^+$-polarized light for Li optical pumping (green curve) as well as without Li optical pumping (blue curve). For the traces in (c), He$^*$ is optically pumped using $\sigma^+$-polarized laser light. The shaded regions indicate the statistical fluctuation of the measured product ion count rate.}
\end{figure}
\noindent The determination of relative magnetic-sublevel populations in He$^*$ is described in detail in Ref.\ \cite{Sixt2021}. By analyzing the polarization of the fluorescence light emitted by the He atoms during laser-optical pumping, we infer a relative population $\eta_{+} \approx 0.92$ in the $M_{J''} = + 1$ magnetic sublevel for excitation using $\sigma^+$-polarized light and $\eta_{-} \approx 0.90$ in the $M_{J''} = - 1$ magnetic sublevel for excitation using $\sigma^-$-polarized light, respectively, which is the same for all valve temperatures in this study. Since $J = S$ in $^4$He, $\eta_{+}$ and $\eta_{-}$  directly represent the relative spin-sublevel populations $\rho_{+1}$ and $\rho_{-1}$, respectively.

The relative magnetic-sublevel populations in Li are determined using absorption spectroscopy via the $F' = 3, 2, 1 \leftarrow F'' = 2$ transitions within the Li D$_2$ transition after optical pumping. This is done in two separate laser frequency scans, one using $\sigma^+$- and one using $\sigma^-$-polarized absorption laser light. In this way, only transitions with a change in the magnetic quantum number corresponding to $\Delta M_F = M_F'-M_F'' = +1$ and $\Delta M_F = -1$ are excited, respectively. Figs.\ \ref{fig:ionsSpectrum} (a) and (b) show representative Li absorption spectra which are obtained after Li optical pumping using $\sigma^+$-polarized light (quarter-wave-plate (QWP) angle of $0^\circ$) and $\sigma^-$-polarized light (QWP angle of $90^\circ$), respectively. In order to infer relative magnetic-sublevel populations from the recorded absorption spectra, calculated Li absorption spectra are fit to the experimental data in a least-squares fitting routine. The relative $F'' = 2, M_{F''}$ sublevel populations are then extracted as fit parameters. Tab.\ \ref{tab:LiMFpopulations} shows the relative $F'' = 2, M_{F''}$ sublevel populations obtained from the absorption spectra in Figs.\ \ref{fig:ionsSpectrum} (a) and (b) using this procedure. More details about the calculation of the absorption spectra and the fitting procedure are given in the appendix. Standard angular momentum algebra is then used to infer the proportional contribution of the $M_S = \pm 1/2$ spin-sublevel populations from the relative $F'' = 2, M_{F''}$ sublevel populations.
\begin{table}[ht!]
	\caption{\label{tab:LiMFpopulations} Relative magnetic-sublevel populations $\rho_{M_{F''}}$ for Li in the 2$^2$S$_{1/2}$, $F'' = 2$ level obtained from the least-squares fitting of calculated absorption spectra to the measured absorption spectra shown in Figs.\ \ref{fig:ionsSpectrum} (a) and (b), respectively. \replaced{Values in parentheses show the statistical uncertainties.}{The uncertainty is $\approx$ 3\,\% for each relative population.}} %
	\begin{tabular}{@{}lcccccccccc@{}}
		\toprule[0.7pt]
		$M_{F''}$ && -2 && -1 && 0 && 1 && 2 \\
		\midrule[0.7pt]
		$\rho_{M_F''}$ (OP Li $\sigma^+$) && 0\added{.00(3)} && 0\added{.00(0)} && 0.02\added{(5)} && 0.25\added{(10)} && 0.73\added{(11)} \\
		$\rho_{M_F''}$ (OP Li $\sigma^-$) && 0.79\added{(3)} && 0.13\added{(4)} && 0.03\added{(3)} && 0.03\added{(3)} && 0.03\added{(3)} \\
		\bottomrule[0.7pt]
	\end{tabular}
\end{table}
\begin{figure}[!htbp]
	\includegraphics{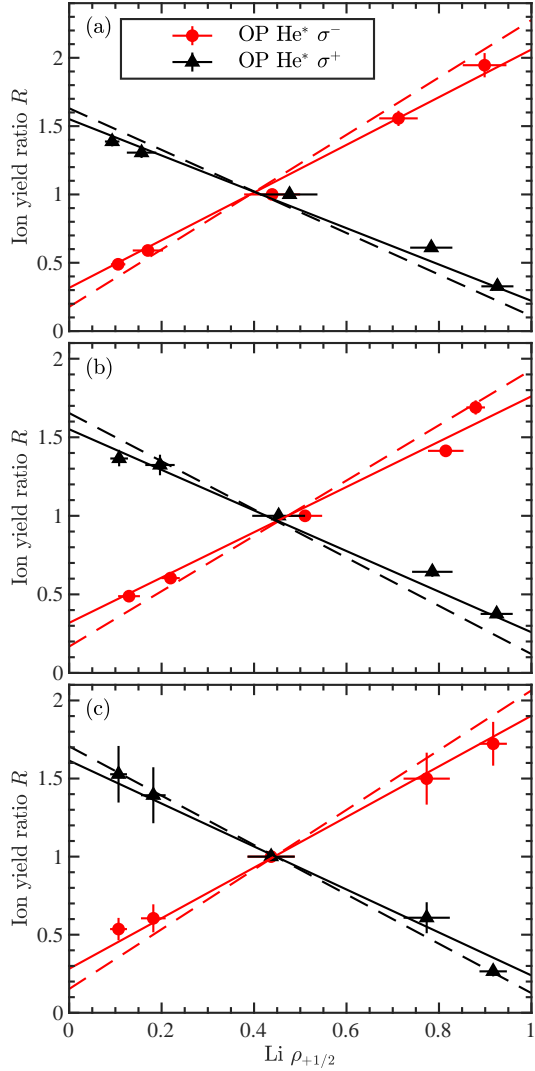}
	\caption{\label{fig:ionRateRatio} Ion yield ratios obtained for chemi-ionization reactions between optically pumped He$^*$ and optically pumped Li as a function of the relative spin-sublevel population in Li at collision energies of (a) $E_\text{coll} = 43.9\unit{meV}$, (b) $E_\text{coll} = 23.3\unit{meV}$ and (c) $E_\text{coll} = 9.6\unit{meV}$. The ion yields are normalized to the equilibrium ion yield obtained from the optical pumping of Li using an equal mixture of $\sigma^+$- and $\sigma^-$-polarized light. The black (red) colors denote determinations in which $\sigma^+$-polarized ($\sigma^-$-polarized) laser light is used for the optical pumping of He$^*$. The results from experimental determinations are shown as colored markers. The solid lines are obtained from fits of the model given in Eq.~\eqref{eq:modelfit} to the corresponding experimental data (see main text). The dashed lines, which correspond to the dashed lines in Fig.\ \ref{fig:heatmap}, show the behavior of the modeled ion yield ratios assuming that $k_4 = 0$. The given experimental uncertainties are statistical only (standard deviation of $2\sigma$).}
\end{figure}

Fig.\ \ref{fig:ionsSpectrum} (c) shows a selection of measured product ion traces for He$^*$-Li chemi-ionization. A strong reduction of the ion count rate is observed when both species are optically pumped by laser light with the same handedness of circular polarization (green curve) compared to the case in which Li is unpolarized (blue curve). Furthermore, if both species are optically pumped by laser light with opposite handedness of circular polarization, there is an increase in the ion count rate (yellow curve). In order to quantify these effects, we normalize the total ion yield for each QWP position used for the optical pumping of Li to the ion yield obtained when the laser polarization for Li optical pumping is composed of an equal mixture of $\sigma^+$ and $\sigma^-$. The latter configuration results in an approximately equal population of the spin sublevels in Li. To extract the rate constant ratio $k_4/k_2$ for each collision energy, Eq.~\eqref{eq:modelfit} is globally fit to the measured ion yield ratios obtained for the optical pumping of He$^*$ using $\sigma^+$- and $\sigma^-$-polarized light, respectively. The equilibrium spin-sublevel population in Li is used as a fit parameter, because systematic asymmetries in the experimental setup lead to a slight spin polarization of the Li atoms in the process of laser optical pumping with an equal mixture of $\sigma^+$- and $\sigma^-$-polarized light. The uncertainties of the measured ion yield ratios are accounted for as weights in the fitting routine. 

The colored markers in Fig.\ \ref{fig:ionRateRatio} show the ion yield ratios obtained in measurements at three different collision energies. The solid lines are ion yield ratios from fits of the model to the experimental data. Model results assuming that $k_4 = 0$ are shown as dashed lines for comparison (cf. dashed lines in Fig.\ \ref{fig:heatmap}). The obtained rate constant ratios $k_4/k_2$ are listed in Tab.\ \ref{tab:temperaturescan}. The uncertainty given for each rate constant ratio $k_4/k_2$ in Tab.\ \ref{tab:temperaturescan} results from the uncertainty of the prepared relative spin-sublevel populations in He$^*$, because these parameters have a direct impact on the best fit value of $k_4/k_2$ without changing the quality of the fit.

From the results given in Tab.\ \ref{tab:temperaturescan}, we infer a small, non-zero contribution of the $^4\Sigma^+$ channel to He$^*$-Li chemi-ionization at thermal collision energies. Only at the lowest experimentally accessible collision energy, the determined rate constant ratio is consistent with zero. The large uncertainty for this determination is a result of the low He$^*$ flux and the resulting large uncertainty in the prepared He$^*$ spin-sublevel population distribution under these conditions.

A breakdown of the spin selection rule for chemi-ionization is usually attributed to the anisotropic coupling between channels of different symmetry. For example, the spin-dipole interaction is reported as the main loss process for homonuclear reactions between ultracold, spin-polarized He$^*$ atoms \cite{Fedichev1996}. In the He$^*$-alkali-atom system, the spin-dipole interaction induces a coupling in between the $^4\Sigma^+$ and $^2\Sigma^+$ channels, so that the $^4\Sigma^+$ channel also receives some $^2\Sigma^+$ character and vice versa.  

Comparing between non-spin-polarized and fully spin-polarized reaction partners, we deduce a reaction suppression factor of $10$ for He$^*$-Li chemi-ionization. This is less than the suppression factor of $>100$ observed for He$^*$-Rb chemi-ionization at collision energies corresponding to ultracold temperatures \cite{Byron2010,Knoop2014,Flores2016,Flores2017}. This deviation may be explained by the increase of the spin-dipole interaction towards higher collision energies which is due to the contribution of higher partial waves to the reaction rate \cite{Fedichev1996}. In fact, the main contribution to the total He$^*$-Li chemi-ionization rate at thermal collision energies is from partial waves with high orbital angular momentum quantum numbers $l$ \cite{Merz1989}. For illustration, Tab.\ \ref{tab:temperaturescan} also provides a list of the maximum orbital angular momentum quantum numbers $l_{\mathrm{max}}$ for which the respective collision energy just exceeds the centrifugal barrier. For the calculation of $l_{\mathrm{max}}$, we use the dispersion coefficients given in Ref.\ \cite{Zhang2012}. Furthermore, effects beyond the Born-Oppenheimer approximation may also be considered for causing a violation of spin conservation in He$^*$-Li chemi-ionization. A quantitative understanding of the observed spin-selection-rule violation will require a comparison with the results from accurate quantum-chemical calculations which are not available to us at this date.
\begin{table}[ht!]
	\caption{\label{tab:temperaturescan} Summary of obtained rate constant ratios $k_4/k_2$ at different collision energies $E_\text{coll}$. For each collision energy, the maximum orbital angular momentum quantum number $l_\text{max}$ is given for which the collision energy just exceeds the centrifugal barrier. The $l_\text{max}$ values are determined using the dispersion coefficients given in Ref.\ \cite{Zhang2012}.}
	\begin{tabular}{@{}lcccc@{}}
		\toprule[0.7pt]
		$E_\text{coll}$ [meV] && $l_\text{max}$ && $k_4/k_2$ $[\%]$ \\
		\midrule[0.7pt]
		$43.9$ && $59$ && $5^{+3}_{-3}$ \\
		$23.3$ && $47$ && $7^{+3}_{-3}$ \\
		$9.6$ && $34$ && $5^{+4}_{-5}$ \\
		\bottomrule[0.7pt]
	\end{tabular}
\end{table}
%
%
\section{\label{sec:conclusions}Conclusions}
\noindent We have demonstrated the control of He$^*$-Li chemi-ionization by magnetic-sublevel preparation of both reaction partners through optical pumping. By adjusting the amount of circular polarization for Li optical pumping, we have been able to tune the chemi-ionization rate by a factor of more than three. The use of state-selective absorption imaging for Li and the analysis of fluorescence emitted during the optical pumping of He$^*$ have allowed us to quantify the relative magnetic-sublevel populations prior to the collision process. In combination with the observed ionization rates, these values have then been used for a quantitative comparison with the results from a model calculation based on the spin angular momentum coupling of the atomic spin states to the quasi-molecular states. From this comparison, we infer a non-vanishing contribution of the $^4\Sigma^+$ channel to He$^*$-Li chemi-ionization at thermal collision energies which is in violation of spin conservation. This effect may be attributed to an increased spin-dipole transition rate at thermal energies compared to the ultracold regime. Thus, at collision energies corresponding to ultracold temperatures, chemi-ionizing collisions between spin-polarized He$^*$ and Li shall be more strongly suppressed, since only partial waves with low or even zero orbital angular momentum contribute to the reaction rate, and a successful co-trapping of He$^*$ and Li is anticipated under these conditions.
%
\begin{acknowledgments}
\noindent The authors thank L. G{\"o}pfert (University of Freiburg), T. Muthu-Arachchige (now at the University of Bonn) and M. Debatin (now at the University of Kassel) for their valuable contributions to the experimental setup. This work was financed by the German Research Foundation (Project No. DU1804/1-1), by the Fonds der Chemischen Industrie (Liebig Fellowship to K.D.), and by the University of Freiburg (Research Innovation Fund).
\end{acknowledgments}

\appendix
\section{Calculation of the Li absorption spectrum}\label{app:absSim}
\noindent For the calculation of the transition frequencies within the Li D$_2$ transition, the eigenenergies of the 2$^2$S$_{1/2}, F'' = 2$ and 2$^2$P$_{3/2}, F' = 1, 2, 3$ levels of Li in the $\left| F, M_F \right\rangle$ basis (eigenstates of $\hat{H}_\text{HFS}$) are calculated using a numerical diagonalization of the Hamiltonian
\begin{align}
	\hat{H} = \hat{H}_\text{HFS} + \hat{H}_\text{Zeeman}.
\end{align}
Here, $\hat{H}_\text{HFS}$ is the hyperfine-structure Hamiltonian and $\hat{H}_\text{Zeeman}$ is the Hamiltonian for the Zeeman interaction. 
The transition strength is given by the product of the relative $\left| F'', M_{F''} \right\rangle$ sublevel population in the 2$^2$S$_{1/2}$ level and the relative magnitude of the corresponding squared transition dipole matrix element. For the calculation of the latter, it has to be considered that, even in the small magnetic fields in our experiment, the hyperfine splitting within the $2^2$P$_{3/2}$ level is small (hyperfine coupling constant $A \approx 3\unit{MHz}$ \cite{Orth1975}) compared to the Zeeman splitting. As a result, the $\left| F', M_{F'} \right\rangle$ states in the 2$^2$P$_{3/2}$ level are no longer good ei\deleted{n}genstates of $\hat{H}$ and they can thus not be used for the calculation of the transition matrix elements. Instead, the transition matrix elements are obtained using \cite{Kaushik2014}
\begin{align}
	\vec{\mu}_{eg} &= \left\langle f',\, m_{f'} \right| \vec{d} \left| f'',\, m_{f''} \right\rangle \\
	&= \sum_{F',\, M_{F'}} \sum_{F'',\, M_{F''}} \left\langle f',\, m_{f'} \middle| F',\, M_{F'} \right\rangle \cdot \left\langle F'',\, M_{F''} \middle| f'',\, m_{f''} \right\rangle \cdot \left\langle F',\, M_{F'} \right| \vec{d} \left| F'',\, M_{F''} \right\rangle,
\end{align}
where $\left| f,\, m_{f} \right\rangle$ are the eigenstates of $\hat{H}$. The transition dipole matrix elements for transitions between the eigenstates of $\hat{H}_\text{HFS}$, $\left\langle F',\, M_{F'} \right| \vec{d} \left| F'',\, M_{F''} \right\rangle$, are calculated using standard procedures \cite{Metcalf1999, Steck2021}.

The absorption spectrum is obtained from a sum of individual Lorentzian functions, where the transition frequency is the center position and the transition strength is the amplitude. The spectrum is fitted to the experimental data in a constrained least-squares manner using a sequential quadratic programming (SQP) algorithm provided by \replaced{MATLAB}{Matlab}. The magnetic field $B_z$, a shift in the energy axis as well as the relative magnetic-sublevel populations $\rho_{M_{F''}}$ for Li in the $2^2$S$_{1/2}, F'' = 2$ level are used as fit parameters. For the determination of the latter, the sum over all relative magnetic-sublevel populations in the $2^2$S$_{1/2}, F'' = 2$ level is constrained to 1.
\section*{Data availability}
\noindent The data that support the findings of this study are available from the corresponding author upon reasonable request.
%

%

\listofchanges 
\end{document}